\documentclass{PoS}

\title{The lattice ghost propagator in Landau gauge \\ 
       up to three loops using \\ 
       Numerical Stochastic Perturbation Theory}

\ShortTitle{The lattice ghost propagator in Landau gauge up to three loops
         using NSPT}

\author{F.~Di~Renzo\\
Universit\`a di Parma \& INFN, Viale Usberti 7/A, I-43100 Parma, Italy\\
E-mail: \email{francesco.direnzo@fis.unipr.it}}

\author{E.-M.~Ilgenfritz\\
Institut f\"ur Theoretische  Physik, 
Ruprecht-Karls-Universit\"at Heidelberg, Philosophenweg 19,  D-69120 Heidelberg, 
Germany \\
E-mail: \email{ilgenfri@physik.hu-berlin.de}}

\author{H.~Perlt\\
Institut f\"ur Theoretische Physik, Universit\"at Leipzig, PF 100 920, D-04009 Leipzig, 
Germany\\
E-mail: \email{Holger.Perlt@itp.uni-leipzig.de}}

\author{\speaker{A.~Schiller}\\
Institut f\"ur Theoretische Physik, Universit\"at Leipzig, PF 100 920, D-04009 Leipzig, 
Germany\\
E-mail: \email{Arwed.Schiller@itp.uni-leipzig.de}}

\author{C.~Torrero\\
Institut f\"ur Theoretische Physik, Universit\"at Regensburg, Universit\"atsstr. 31, 
\\
D-93053 Regensburg, Germany\\
E-mail: \email{christian.torrero@physik.uni-regensburg.de}}

\abstract
{We complete our high-accuracy studies of the lattice ghost propagator in Landau gauge
in Numerical Stochastic Perturbation Theory up to three loops.
We present a systematic strategy which allows
to extract with sufficient precision the non-logarithmic parts of logarithmically divergent
quantities as a function of the propagator momentum squared
in the infinite-volume and $a\to 0$ limits.
We find accurate coincidence with the one-loop result for the ghost self-energy
known from standard Lattice Perturbation Theory and improve our previous estimate
for the two-loop constant contribution to
the ghost self-energy in Landau gauge.
Our results for the perturbative ghost propagator are compared with Monte Carlo
measurements of the ghost propagator performed by the Berlin Humboldt university
group which has used the exponential relation between
potentials and gauge links.
}

\FullConference{The XXVII International Symposium on Lattice Field Theory
\\
		 July 26-31, 2009\\
		 Peking University, Beijing, China}

\begin{document}

\section{NSPT and Langevin equation}

It is known that standard diagrammatic Lattice Perturbation Theory (LPT) becomes very
complicated when studying higher orders of typical physical quantities as renormalization
factors.

As an alternative, Numerical Stochastic Perturbation Theory (NSPT) 
{\small (see e.g.~\cite{DiRenzo:2004ge})} is a powerful tool to study higher-loop contributions 
in LPT: thanks to it, higher-loop results are in fact obtained without computing vast numbers of
Feynman diagrams.
Several applications of NSPT have been reported over the last years, for some 
latest  developments see
the additional contributions of F. Di Renzo, M. Brambilla,  C. Torrero and H. Perlt 
to this conference.

Here we extend our results reported earlier~\cite{DiRenzo:2008ir,DiRenzo:2008nv} 
and study the three-loop ghost propagator in Landau gauge
to make predictions for standard diagrammatic LPT and 
compare with non-perturbative calculations.

We use the lattice Langevin equation with stochastic time $t$
\begin{eqnarray}
  \frac{\partial}{\partial t} U_{x,\mu}(t;\eta) = {\rm{i}}  \left(
  \nabla_{x,\mu} S_G[U]+ \eta_{x,\mu}(t)
  \right) \;
  U_{x,\mu}(t;\eta)\,,
\end{eqnarray}
where $\eta$ is  Gaussian random noise, $S_G$ the gauge action and
$\nabla_{x,\mu}$ the left Lie derivative within the gauge group.
Discretizing the time $t= n \tau$, the
equation is integrated numerically in the Euler scheme  by iteration:
\begin{equation}
  U_{x,\mu}(n+1; \eta)= {\rm{exp}}(-F_{x,\mu}[U,\eta])\; U_{x,\mu}(n; \eta)
  \label{eq:update_step}
\end{equation}
with the force
\begin{equation}
  F_{x,\mu}[U, \eta]= {\rm{i}}(\tau \nabla_{x,\mu} S_G[U] + 
  \sqrt{\tau} \, \eta_{x,\mu}) \,.
  \label{eq:force}
\end{equation}

Rescaling $\varepsilon = \beta \tau$ and expanding the gauge links
($g \propto \beta^{-1/2}$)
\begin{eqnarray}
  U_{x,\mu} \to 1 + \sum_{l>0} \beta^{-l/2} U_{x,\mu}^{(l)}\,,
\end{eqnarray}
the Langevin equation at finite time step $\varepsilon$ turns into a
system of updates for each perturbative order {$U_{x,\mu}^{(l)}$}.
The algebra-valued gauge potentials $A_{x,\mu}$
are related to the gauge lattice link fields $U_{x,\mu}$ by
\begin{equation}
  A_{x,\mu}= \log U_{x,\mu} \, .
  \label{eq:Adef}
\end{equation}
Their expansion is given in the form 
\begin{equation}
  A_{x,\mu} \to \sum_{l>0} \beta^{-l/2} A_{x,\mu}^{(l)} 
  \label{eq:A-expansion}
\end{equation}
and is used to enforce unitarity to all orders in $1/\sqrt{\beta}$.

Each simultaneous Langevin update  is augmented
by a stochastic gauge-fixing step and by subtracting zero modes from $A^{(l)}$.
{}From the  resulting fields the Green functions of interest 
can be numerically constructed order by order.

To measure gauge-dependent quantities, exact gauge fixing is needed.
We use the Landau gauge which is reached by iterative Fourier-accelerated 
gauge transformations~\cite{Davies:1987vs}.

\section{The ghost propagator in perturbation theory}

\subsection{The ghost propagator in NSPT}

It is known that the ghost propagator is defined from the inverse of the 
Faddeev-Popov (FP) operator
$M$. In Landau gauge this operator is constructed by using the lattice covariant  $D(U)$
and left partial derivatives, $M= - \nabla \cdot D(U)$. Following~\cite{Rothe:1997kp}
we use here a definition of $M$ which is most suitable for NSPT. 

We introduce the physical lattice momenta 
$
  \hat p_{\mu}(k_{\mu}) = \frac{2}{a} \sin\left(\frac{\pi
 k_{\mu}}{L_{\mu}}\right)= \frac{2}{a} \sin\left(\frac{ a p_\mu}{2}\right) 
$
and define the color diagonal propagator in momentum space
as the color trace in the adjoint representation ($N_c=3$)
\begin{equation}
  G( p(k)  )=
  \frac{1}{N_c^2-1} \left\langle {\rm{Tr_{adj}}}~M^{-1}(k)\right\rangle_U \,.
  \label{eq:G-definition}
\end{equation}
In (\ref{eq:G-definition}) $M^{-1}(k)$ is the Fourier transform of the inverse
FP operator in lattice coordinate space.

A perturbative expansion is based on the mapping 
$A_{x,\mu}^{(l)} \,\rightarrow \, M^{(l)} \,\rightarrow \, \left[M^{-1}\right]^{(l)}$ 
which allows to calculate the inverse FP operator in NSPT recursively
(i.e., to any finite order inverting the matrix $M$ results in a closed
 form)
\begin{eqnarray}
  \left[M^{-1}\right]^{(0)} = \left[M^{(0)}\right]^{-1} \,, \quad
  \left[M^{-1}\right]^{(l)} = -\left[M^{(0)}\right]^{-1}\sum_{j=0}^{l-1} M^{(l-j)}
  \left[M^{-1}\right]^{(j)}\,.
\end{eqnarray}

The momentum space ghost propagator is obtained by sandwiching
$\left[M^{-1}\right]^{(l)}$ between plane-wave vectors.
The propagator has to be computed from scratch for each chosen momentum tuple
$(k_1,k_2,k_3,k_4)$ and different colors of the plane wave. Since we are interested in finding 
the momentum behavior as good as possible,
these measurements become relatively expensive.

Multiplying the measured lattice momentum ghost propagator either with $(a p)^2$ or
$\hat p^2$, two forms of the so-called ghost dressing function are defined:
\begin{eqnarray}
  J^{(l)} = (a p)^2 \; G^{(l)} \,, \quad
  \hat J^{(l)} = \hat p^2 \; G^{(l)} \,.
\end{eqnarray}

The perturbative construction of $M$ in terms of $A$
differs from the definition adopted in most Monte Carlo calculations
where a linear relation between the gauge links and gauge potentials is used.

\subsection{The ghost propagator in standard LPT}

In the RI'-MOM scheme, the renormalized ghost dressing function $J^{\rm RI'}$ is
defined as
\begin{eqnarray}
  J^{\rm RI'}( p, \mu , \alpha_{\rm RI'}) =
  \frac{J(a, p, \alpha_{\rm RI'})}{Z_{\rm gh}(a,\mu,\alpha_{\rm RI'})} 
\end{eqnarray}
with the renormalization condition
$J^{\rm RI'}( p, \mu , \alpha_{\rm RI'})|_{p^2=\mu^2} = 1$.
Therefore, the ghost dressing function $J(a,p,\alpha_{\rm RI'})$ is just the ghost
wave function renormalization constant $Z(a,\mu,\alpha_{\rm RI'})$ at $\mu^2=q^2$.

We represent the expansion of the bare $J(a,p, \alpha_{\rm RI'})$ to n-loop order by
\begin{eqnarray}
  J^{\rm{n-loop}}(a,p, \alpha_{\rm RI'})&=&1+
  \sum_{i=1}^{n} \alpha_{\rm RI'}^i \,\sum_{k=0}^i\, z^{\rm RI'}_{i,k}\,
  \left(\frac{1}{2} {\cal L} \right)^{\!\!k} \,, \ \ \ \ \ \ {\cal L}=\log (ap)^2 \,.
\end{eqnarray}
The leading log coefficients $z^{\rm RI'}_{i,i}$ coincide with continuum
perturbation theory (PT), the subleading log coefficients $z^{\rm RI'}_{i,k}|_{i>k>0}$
have to be determined both from  continuum PT and LPT in the used scheme,
$z^{\rm RI'}_{i,0}$ has to be found in LPT.
Restricting ourselves in these proceedings to two-loop order in Landau gauge and to 
quenched approximation (compare e.g.~\cite{Gracey:2003yr}), 
we have 
\begin{eqnarray}
  z^{\rm  RI'}_{1,1}= - \frac{9}{2} \,, \quad
  z^{\rm  RI'}_{2,2}= - \frac{315}{8}  \,, \quad
  z^{\rm RI'}_{2,1}= - \frac{2439}{24} +
  \frac{35}{2} \, z^{\rm RI'}_{1,0}   \,.
\end{eqnarray}
The coefficient $z^{\rm RI'}_{1,0}=13.8257$  has been calculated in~\cite{Kawai:1980ja},
a first rough estimate of
$z^{\rm  RI'}_{2,0}$ has been presented by us at Lattice08~\cite{DiRenzo:2008ir}.

{}From { $\alpha_{\rm RI'}= \alpha_0 + \left(-22\log (a\mu) + 73.9355 \right)
\alpha_0^2 + \dots$}~\cite{Hasenfratz:1980kn,Luscher:1995np},
with the bare coupling $\alpha_0=3/(8 \pi^2 \beta)$, we get for the
dressing function
\begin{eqnarray}
  J^{\rm{2-loop}}(a,p, \beta) = 1 + \frac{1}{\beta} \left(J_{1,1} {\cal L}
  + {J_{1,0}} \right) +
  \frac{1}{\beta^2}\left(J_{2,2}{\cal L}^2 +J_{2,1} {\cal L} + {J_{2,0}}\right)
\end{eqnarray}
with
\begin{eqnarray}
   J_{1,1}=-0.0854897 \,, \ \  J_{1,0}=  0.525314\,,
   \ \ J_{2,2}= 0.0215195 \,,  \ \ J_{2,1}= -0.358423 \,.
\end{eqnarray}
Among others we have to confirm $\ \! J_{1,0}\ \!$ from standard LPT and want to find a precise
number for $\ \!J_{2,0}\ \!$ in Landau gauge quenched QCD using NSPT.

\section{Results}

\subsection{Practice of measurements}

\begin{figure}[!htb]
  \begin{center}
     \begin{tabular}{cc}
        \includegraphics[scale=0.54,clip=true]
         {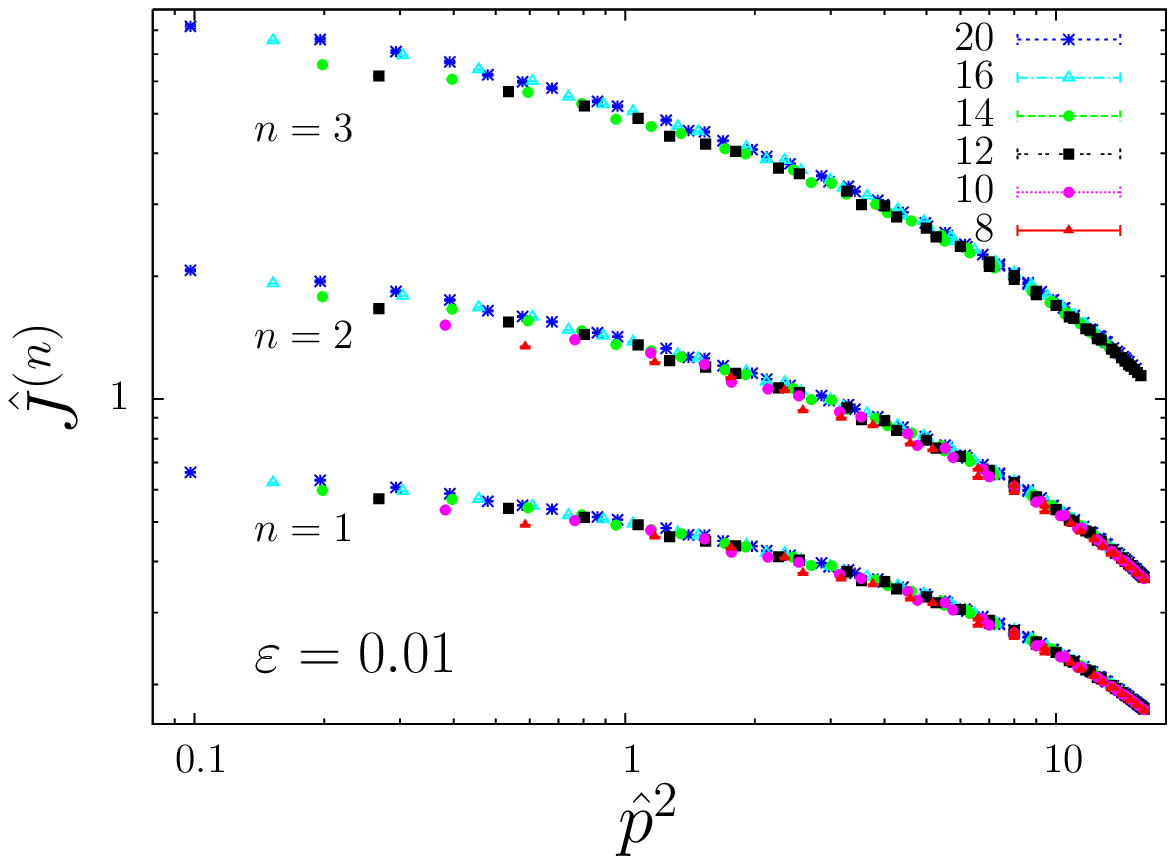}
        &
        \includegraphics[scale=0.54,clip=true]
         {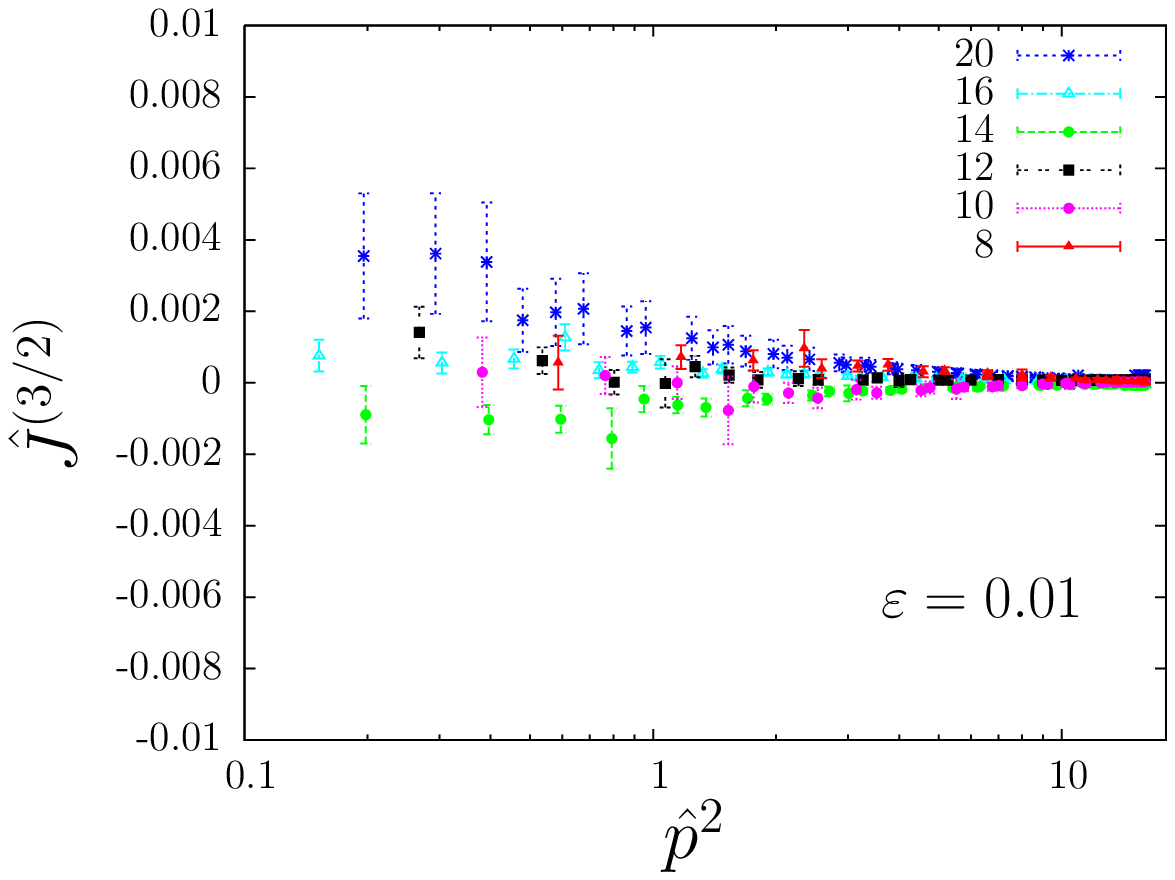}
     \end{tabular}
  \end{center}
  \vspace{-5mm}
  \caption{{Measured ghost dressing function $\hat{J}$ vs. $\hat p^2$
           for all inequivalent lattice momentum  4-tuples near diagonal for
           $\ \!L=8,10,12,14,16,20\ \!$ and $\ \!\varepsilon=0.01\ \!$.
           Left: The one-loop ($\beta^{-1}$), two-loop ($\beta^{-2}$) and three-loop 
           ($\beta^{-3}$) contributions,
           right: the vanishing ($\propto \beta^{-3/2}$) contribution.}}
  \label{fig:1}
\end{figure}

Very precise measurements for different lattice sizes and different Langevin steps
$\varepsilon$ are needed. Typically we have of the order of 1000 measurements
for each momentum tuple.
Already at finite $\varepsilon$ the non-integer $n=l/2$ (no-loop) contributions
to the dressing function have to become negligible compared to the
neighboring loop contributions.
Examples of the dressing function $\hat{J}$  for $n=1,2,3$ and $n=3/2$ 
vs. $\hat p^2$ at different volumes and $\varepsilon=0.01$ are shown in Figure~\ref{fig:1}.
\newpage

\begin{figure}[!htb]
  \begin{center}
     \begin{tabular}{cc}
        \includegraphics[scale=0.54,clip=true]
         {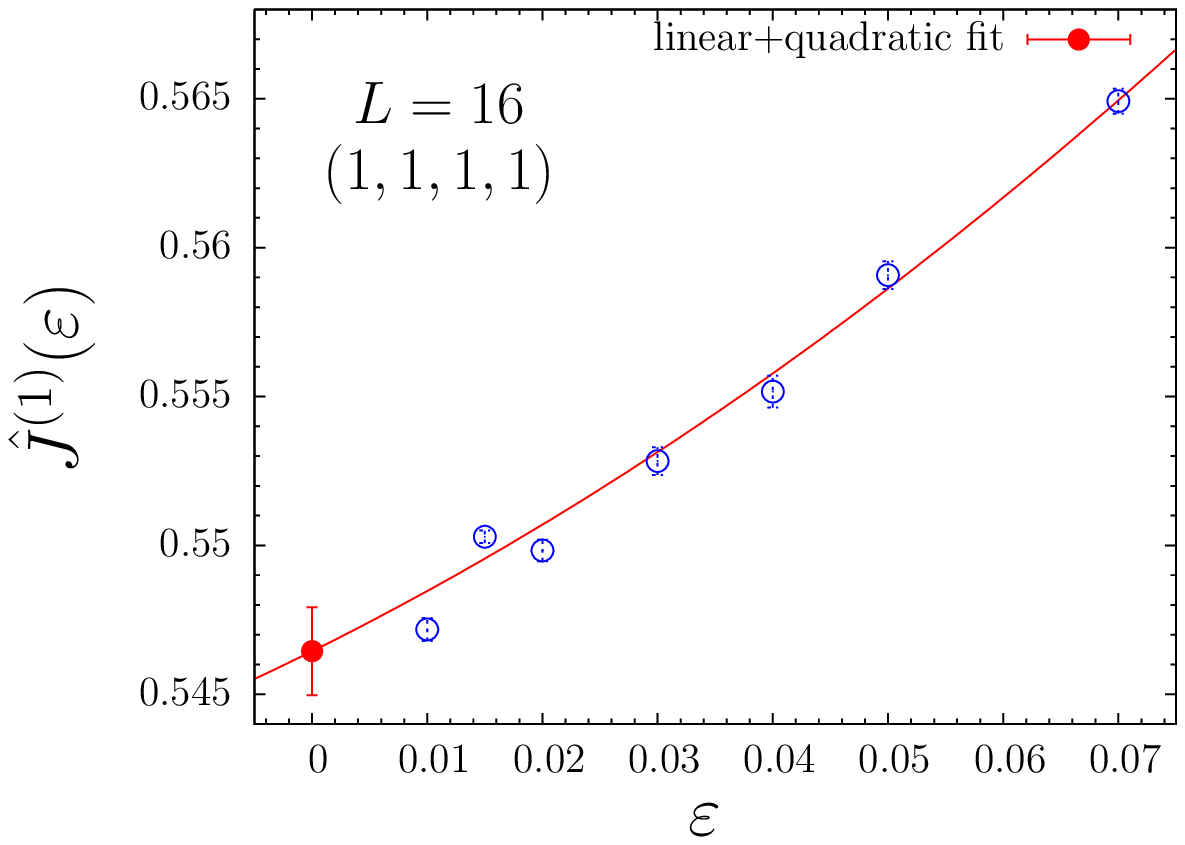}
        &
        \includegraphics[scale=0.54,clip=true]
         {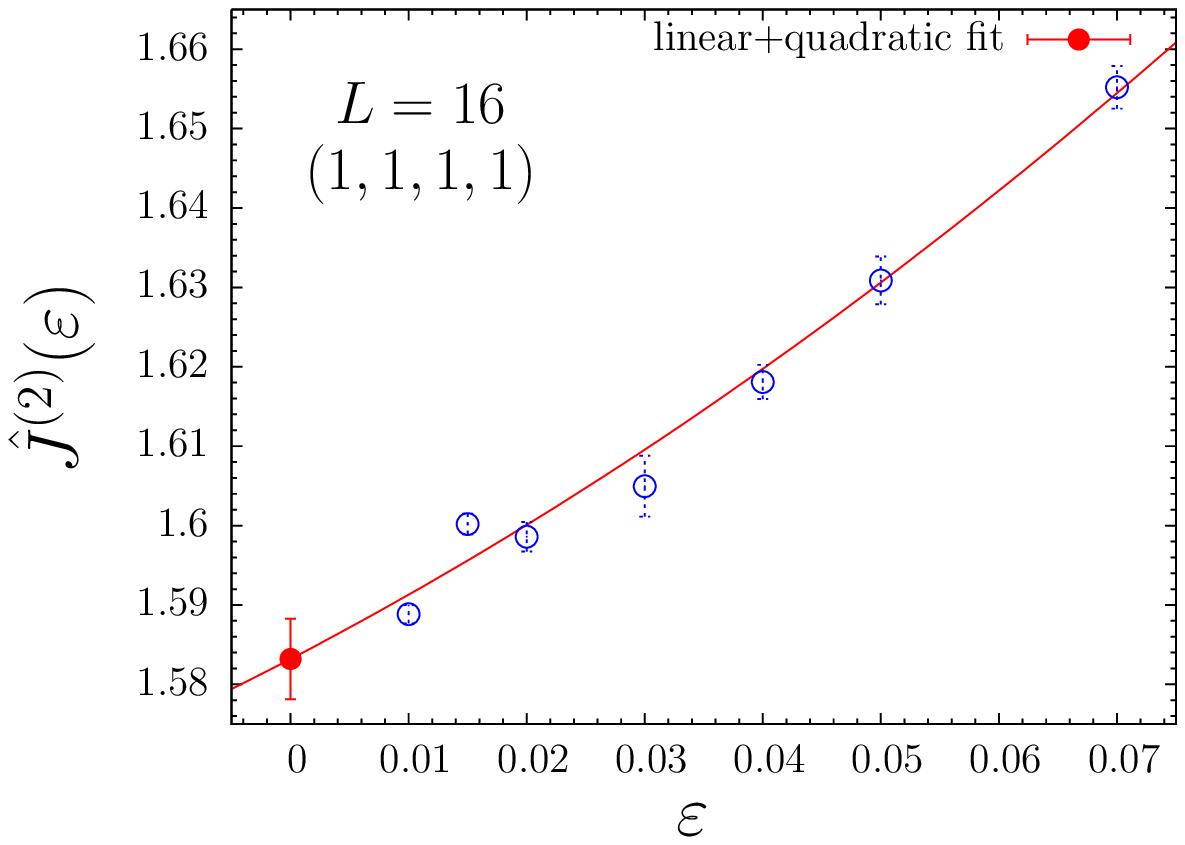}
     \end{tabular}
  \end{center}
  \vspace{-5mm}
  \caption{Extrapolation to $\varepsilon=0$
           of one- (left) and two-loop (right) ghost dressing function for
           lattice size $16^4$ and momentum tuple  $(1,1,1,1)$: the fitting function 
           contains both a linear and a quadratic term in $\varepsilon$.}
  \label{fig:2}
\end{figure}

We have to take the zero Langevin step limit $\varepsilon \to 0$ for each 4-tuple 
{{$(k_1,k_2,k_3,k_4)$}} 
from the available finite $\varepsilon$ measurements
at fixed lattice size. This is shown for a particular momentum 
tuple at $L=16$ in Figure~\ref{fig:2}.
In order to make contact with standard LPT, the limits 
$\ \!L \to \infty\ \!$ and $\ \!a p\to 0\ \!$ have
to be performed additionally.

\subsection{Fitting logarithmic quantities on finite L}
\vspace{-1mm}

The dressing function $J$ at zero Langevin step still suffers from finite
$O(ap)$ and $O(pL)$ corrections.
Our aim is to extract the finite constants in the power expansion of the
lattice ghost dressing function $\ \!J_{i,0}\ \!$ with very high accuracy.
In~\cite{Di Renzo:2006wd} it was pointed out that finite-size effects can be large when
an anomalous dimension comes into play. Having at hand a variety of lattice sizes, 
we address a careful assessment of these effects 
(the main ideas entering the procedure can be found in~\cite{DiRenzo:2007qf}).
Without entering into details, we summarize that strategy of fitting simultaneously
$O(ap)$ and $O(pL)$ corrections together using several lattice sizes.

First we subtract all logarithmic pieces (supposed to be universal and known)
from the dressing function for each momentum tuple and all lattice sizes. 
Next we select a range in 
$(ap)^2= \sum (ap_\mu)^2$, $p_\mu= k_\mu (2 \pi/L)$ with
$p^2_{\rm min} <p^2 <p^2_{\rm max}$.
Within that range we identify a set $S$ of momentum tuples $(k_1,k_2,k_2,k_4)$ 
which is common to all chosen lattice sizes. 
The data in that set are assumed to have the same $pL$ effects.
Since finite-volume effects decrease with increasing momentum squared, 
we  choose as reference fitting point -- for an assumed behavior at 
$L=\infty$ -- the data point at $p^2 \approx p^2_{\rm max}$ from the largest 
lattice size at our disposal.

Next we perform a non-linear fit using 
all data points of different $L$ from that set $S$ and the reference point
correcting for finite size ($C_m$, no functional form)
and assuming a functional behavior for $H(4)$ ($p^n=\sum_\mu p_\mu^n$):
\vspace{-1mm}
\begin{eqnarray}
  &&J_{i,0}(k_1,k_2,k_3,k_4, p^2)= {J_{i,0}} +a^2 \left(\tilde{{\alpha}} \, p^2 +{\tilde{\gamma}}
  \, \frac{p^4}{p^2}\right) +
  a^4\left(  {\tilde{\beta}} \, (p^2)^2
  + {\tilde{\eta}} \, p^4 + {\tilde{\sigma}} \, \frac{p^6}{p^2}\right)+\nonumber
  \\
  &&
  + \sum_{m \in S} {C_m} \, \delta[m,\{k_1,k_2,k_3,k_4\}]  \,, \quad i=1,2,\dots
  \label{fitform}
\end{eqnarray}

\begin{figure}[!htb]
  \begin{center}
     \includegraphics[scale=0.68,clip=true]
      {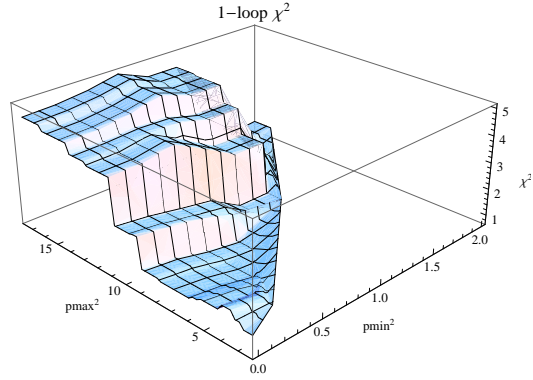}
  \end{center}
  \vspace{-8mm}
  \caption{$\chi^2$-behavior of the non-linear fit for the one-loop
           dressing function in possible lattice momentum windows.} 
  \label{fig:3}
\end{figure}

Finally we vary the momentum squared window and find an optimal $\chi^2$ 
region which allows  us to find the 'best' $J_{i,0}$.
In Figure~\ref{fig:3} 
we show such a $\chi^2$-behavior for a non-linear fit 
to the one-loop dressing function.

\begin{figure}[!htb]
  \begin{center}
    \begin{tabular}{cc}
       \includegraphics[scale=0.57,clip=true]
       {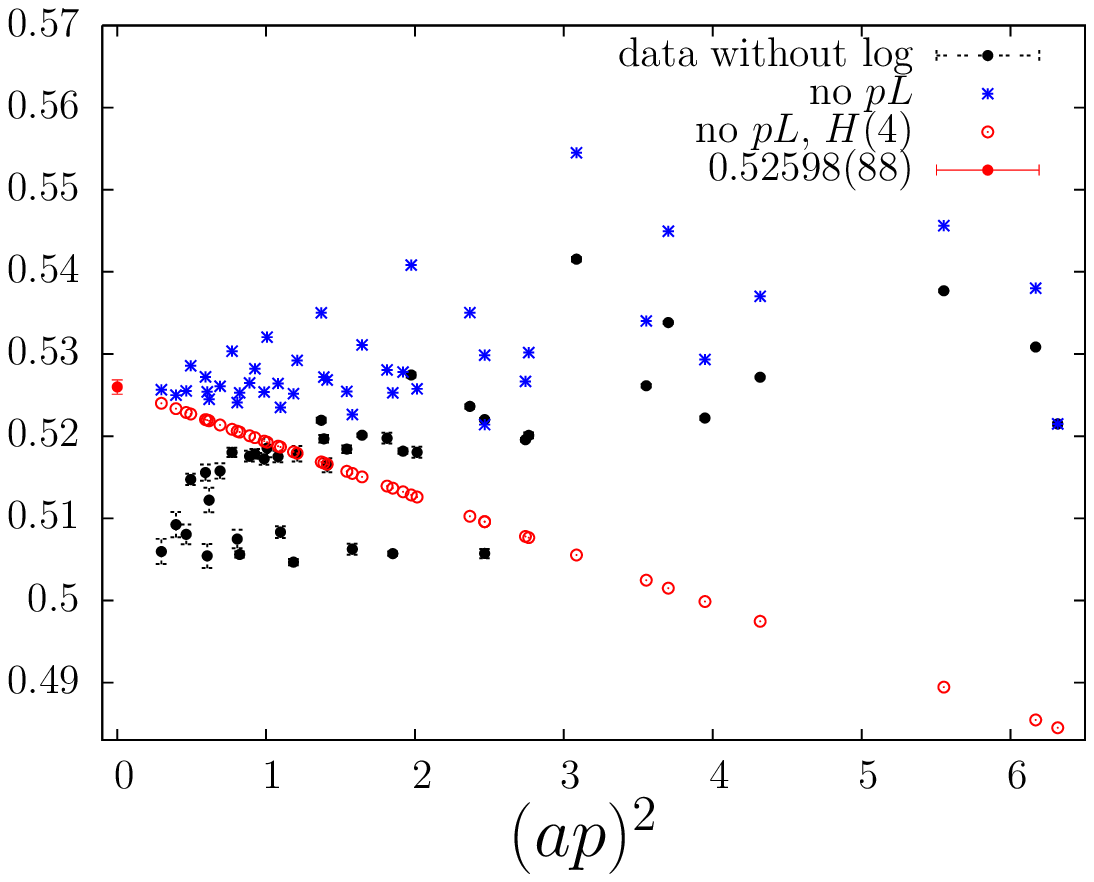}
       &
       \includegraphics[scale=0.57,clip=true]
       {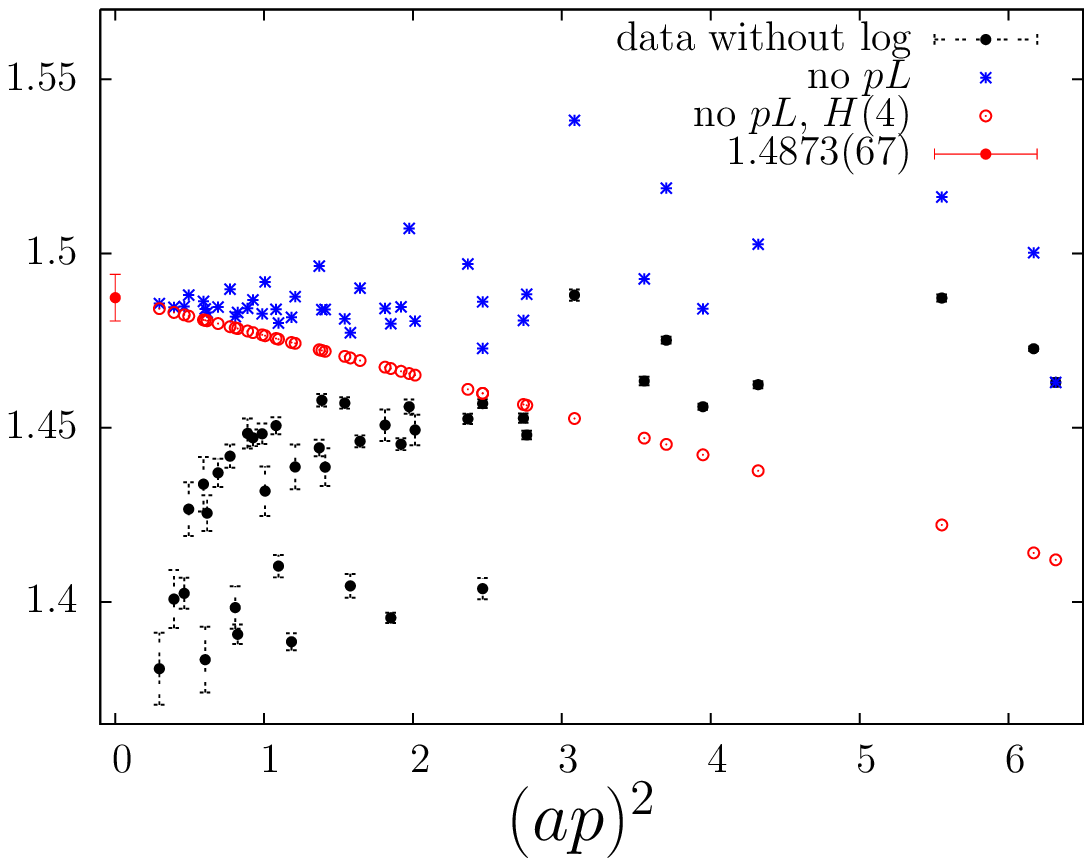}
    \end{tabular}
  \end{center}
  \vspace{-5mm}
  \caption{$L=8,\dots,20$, $S=7$;
           black filled circles: log-subtracted data from allowed tuples;
           blue stars: $pL$ effects removed;
           red open circles: $C_i=0$, $H(4)$ effects removed. 
           Left: one-loop order; right: two-loop order.}
  \label{fig:4}
\end{figure}

An example of a combined fit at low $\chi^2$ is shown in Figure~\ref{fig:4}
for the first and the second loop as a function of $(a p)^2$.
We observe that the numerical data at $\varepsilon=0$  from the 
chosen set -- with all logarithmic contributions subtracted --
scatter significantly (black filled circles). Switching off the $O(p L)$ 
corrections ($C_m=0$ in the fit form~(\ref{fitform})),
the blue stars line up in 'rows' according to 
the different hypercubic invariants at infinite lattice volume. 
The reference point (here the rightmost point) is of course unchanged.
Finally, after removing also the non-rotational hypercubic effects
(leaving only $\alpha \ne 0$ in (\ref{fitform})),
we obtain a smooth (almost linear) curve  formed by the red open circles which 
directly points towards the fitting constant $J_{i,0}$ in the zero lattice spacing limit. 

To make a more realistic  estimate of $J_{i,0}$, we have
collected fit results from five different sets $S$ with minimal $\chi^2$.
{}From these sets we obtain the
following (preliminary) constants in Landau gauge ($J_{1,0}^{\rm{exact}}=0.525314$)
\begin{eqnarray}
    J_{1,0}=0.52520(46) \,, \quad     J_{2,0}= 1.489(4) \,.
\end{eqnarray}
The two-loop constant can be transformed into the RI'-MOM scheme in that gauge. 
Our prediction for that small contribution is
$ z_{2,0}^{\rm RI'}= 9.2 (2.7)$.

Using the found $\ \!J_{2,0}\ \!$ as input for the non-leading log contribution to three loops,
we are in the position to estimate $J_{3,0}$ as well. This analysis
is in progress.

\subsection{Comparison to Monte Carlo data}

Using the $A=-{\rm log} \,  U$ definition as in NSPT, the Berlin Humboldt university group
has produced Monte Carlo results for the ghost and gluon propagator in 
Landau gauge and different gauge couplings~\cite{Menz}.
Since it is assumed that non-perturbative contributions dominate mainly the infrared, 
it is of interest to compare directly the perturbative ghost dressing function obtained in NSPT  
with its Monte Carlo counterpart for each common momentum tuple.

We calculate the perturbative dressing function at a given lattice volume 
summed up to loop order $n_{\rm {max}}$ for a given lattice coupling $\beta$ as follows:
\begin{equation}
  \hat J=\sum_{n=1}^{n_{\rm {max}}} \frac{1}{\beta^n}
  \, \hat J^{(n)}\,.
\end{equation}

\begin{figure}[!htb]
  \begin{center}
     \includegraphics[scale=0.65,clip=true]
     {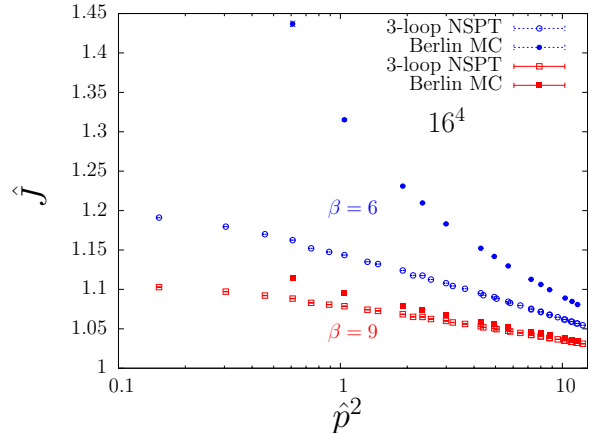}
  \end{center}
  \vspace{-5mm}
  \caption{Three-loop NSPT of the ghost dressing function in comparison to Monte Carlo
           at two $\beta$ values.}
  \label{fig:5}
\end{figure}

In Figure~\ref{fig:5} 
we compare the perturbative ghost dressing function
at lattice size $L=16$ to Monte Carlo data at two different $\beta$ values.
We observe that in not less than three-loop accuracy the perturbative
ghost propagator at larger $\beta$ is approximately able to describe
the full two-point function in the large momentum squared region. 

\begin{figure}[!htb]
  \begin{center}
    \begin{tabular}{cc}
      \includegraphics[scale=0.57,clip=true]
      {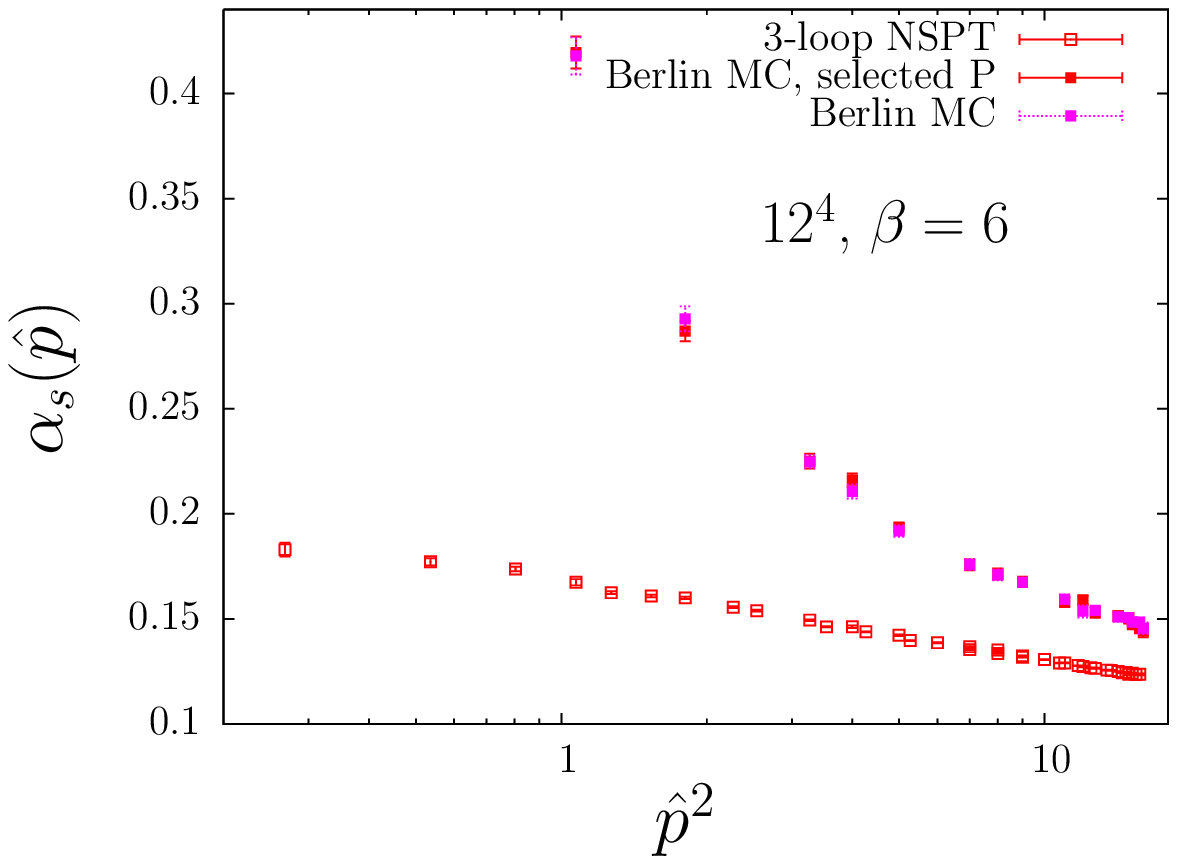}
      &
      \includegraphics[scale=0.57,clip=true]
      {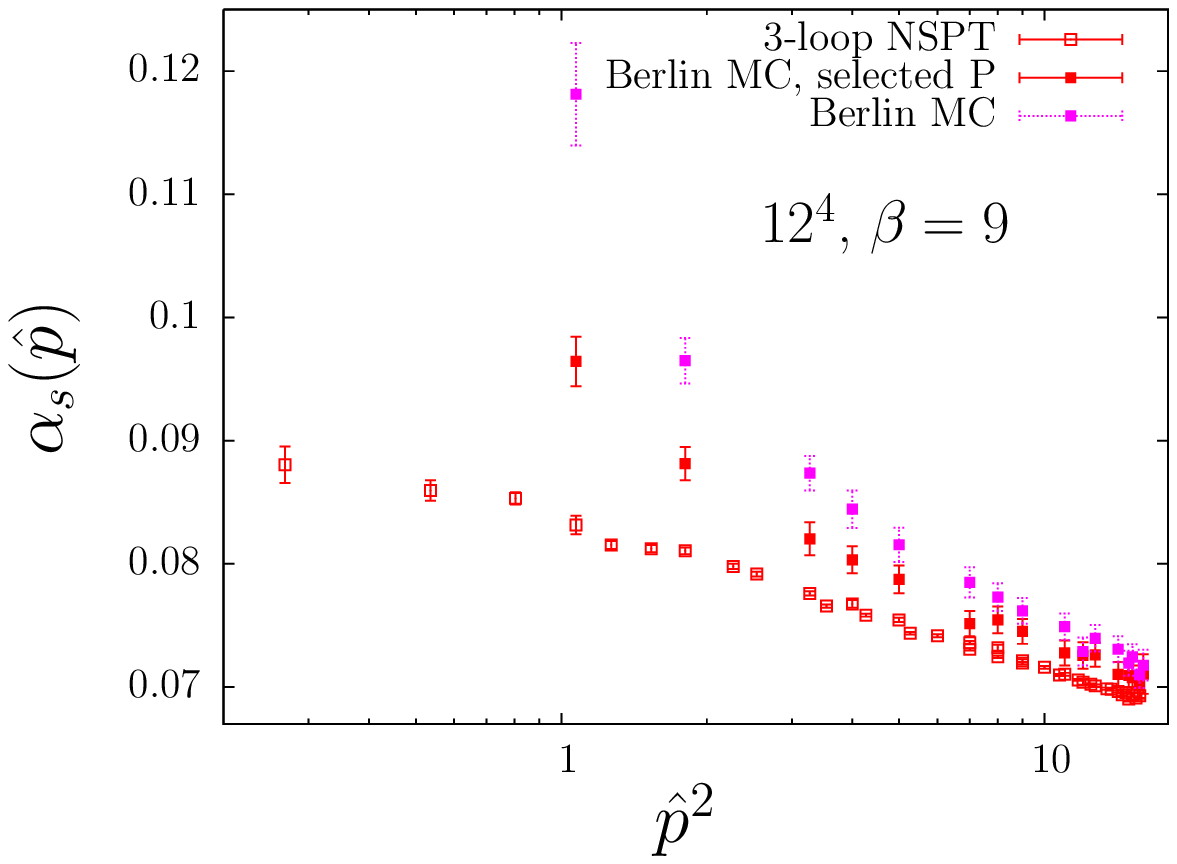}
    \end{tabular}
  \end{center}
  \vspace{-5mm}
  \caption{Three-loop NSPT of $\alpha_s$ in comparison to Monte Carlo
           (partly with specially chosen Polyakov sectors) for two $\beta$ values.}
  \label{fig:6}
\end{figure}

The situation becomes even worse in comparison with Monte Carlo
when trying to define a perturbative running coupling from
\begin{eqnarray}
  \alpha_s^{\rm 3-loop} (\hat p,\beta)= \frac{6}{4 \pi \beta} 
  \hat J(\beta)^2 \,  \hat G(\beta)\,. 
\end{eqnarray}
Here $\hat G$ is the gluon dressing function 
(see~\cite{Ilgenfritz:2007qj,DiRenzo:2008nv}, used in NSPT in the same accuracy).
This is demonstrated in Figure~\ref{fig:6}. 
So more loops would be necessary to find a satisfactory agreement with the 
non-perturbative data at largest lattice momenta.

\section{Summary}

We have presented a detailed perturbative calculation of the lattice ghost propagator
in Landau gauge using NSPT.
The one-loop constant $\ \!J_{1,0}\ \!$ perfectly agrees with known $V \to \infty$ result.
The two-loop constant $\ \!J_{2,0}\ \!$ is determined with good accuracy for the first time.
We have performed a very careful analysis of all necessary limits.
A technique to simultaneously deal with both $O(ap)$ and $O(pL)$ corrections 
is described in some detail.
A comparison with Monte Carlo data of the ghost propagator and the running coupling shows
that additional loops are needed to better describe the asymptotically prevailing
perturbative tail at large lattice momenta.

\end{document}